\documentclass[twocolumn,showpacs,preprintnumbers,amsmath,amssymb]{revtex4}

\usepackage{graphicx}
\usepackage{dcolumn}
\usepackage{bm}

\begin{document}


\title{Detecting Cosmic Gravitational-wave Background from Super-heavy Cosmic Strings with LISA}

\author{Alf Tang}
\affiliation{Blackett Laboratory, Imperial College London, Prince Consort Road, London, SW7 2AZ}
\email{alf.tang@nspo.narl.org.tw}

\author{Timothy J. Sumner}
\affiliation{Blackett Laboratory, Imperial College London, Prince Consort Road, London, SW7 2AZ}
\email{t.sumner@imperial.ac.uk}

\date{\today}

\begin{abstract}
Although cosmic string scenario for galaxy formation is disfavored by CMB data, it is of great interest in the generation of cosmic gravitational-wave background.  This research aims to develop an algorithm to extract cosmic gravitational-wave background produced by cosmic strings from the LISA data stream, and apply the algorithm to the simulated data stream containing the background produced by cosmic strings with various strength to study the detection threshold for this source.  For 1-yr observation, It is found that the detection threshold of $G\mu$ is $3.12\times 10^{-16}$ in the standard scenario.  
In the case that $p$ and $\epsilon$ are adjustable, the detectable region in parameter space is defined by 
$(G\mu)^{2/3}p^{-1}\epsilon^{-1/3}>4.6\times 10^{-11}$.  

\end{abstract}

\maketitle

\section{\label{sec:level1}Introduction}

The cosmic gravitational-wave background (CGB) is the superposition of relic gravitational
waves generated in the very early universe.  Several scenarios might produce CGB have been introduced \cite{CGB}, including inflation \cite{CGB,Inflation,Inflation2,Inflation3,Inflation4,Inflation5,Inflation6,Inflation7}, vacuum-bubbles colliding\cite{VacuumBubble,VacuumBubble2,VacuumBubble3,VacuumBubble4,VacuumBubble5,VacuumBubble6}, 
and cosmic strings \cite{CosmicString,CosmicString2,CosmicString3}.  
Inflation scenario, favored by the Cosmic Microwave Background Radiation (CMBR) data, 
introduces perturbations in all fields, so it is expected the CGB generated by inflation would exist.  
However, the strain amplitude of CGB given by this scenario is below the LISA sensitivity.  
As for vacuum-bubble colliding and cosmic string scenarios, they are the product of the phase transition.  
Though both scenarios could create detectable CGB by LISA, in this paper we will focus on the latter case.

Contrary to localized sources such as binaries, 
the cosmological sources are randomly distributed across the sky and the signals are incoherent, causing an unresolved average of the variation of arm-length and producing a continuum which is entangled 
with instrumental noise.  Moreover, the existence of cosmic gravitational-wave backgrounds is still unclear.  
Therefore, the algorithm used to extract CGB is demanded not only being able to separate CGB from instrumental noise, but also able to decide whether specific kinds of the backgrounds appear in a data set.

The proof mass will be drifting away from its free-falling trajectory due to persisting perturbation by random acceleration.
Despite that the drift itself cannot be modeled because its behavior is analogous to random walk, 
the trend of the drift can be modeled by
a quadratic function of time $at^{2} + bt + c$ where a, b, c are parameters need to be fitted. 
Having removed the trend from the time series data and then Fourier transformed the residual, 
one can obtain the noise spectrum in the frequency domain.  
The low frequency part of the spectrum is regarded as inherent acceleration noise.  
In case that CGB existing in the data, inaccurate estimate might be induced in
data analysis if the trend and the background are sequentially extracted from the data stream.  
To avoid the bias, the analysis shall be performed entirely either in the time domain, or in the frequency domain.  
The difficulty to deal with the both simultaneously was that the stochastic nature of CGB obstructs the extraction from the time domain, but the function of the power spectrum of the trend in the frequency domain was unknown.  
It is natural to extract the CGB in the frequency domain because the power spectrum with respect to a specific kind of background has its own feature, making itself able to be distinguished from others.  
To successfully separate the background and the trend in a data set, we have derived the function of the power
spectrum of the trend induced from the random acceleration in the frequency domain \cite{Tang1} 
so that the trend can be estimated simultaneously with the background.

Data analysis is a time-consuming task.  In general, the time required for completing estimation with brute force 
is roughly exponential increasing with the number of parameter.  
Suppose one has a model with 10 parameters.  
Even if one just has ten trial values for each parameter, 10 billion parameter sets are produced in order to 
find the best estimate.  
To solve this, the Metropolis algorithm \cite{MCMC} has been adapted in our algorithm \cite{Tang1}, 
which is a kind of Markov Chain Monte Carlo (MCMC) method to estimate parameters.  
The time used for parameter estimation with a MCMC is roughly proportional to the number of parameters.  
Furthermore, we have applied the simulated annealing method to speed up the searching of the Markov Chain \cite{Simulated-Annealing,Cooling-Schedule,Cooling-Schedule2}.  
The Gelman \& Rubin method was chosen as a diagnostic for the convergence of the chains to
confirm the robustness of the estimation \cite{Convergence-Review}.  
In our previous work \cite{Tang1}, the algorithm has been constructed based on Bayesian statistics with the three techniques to perform parameter estimation.  
Here, we will extend the capability of the algorithm to perform model selection task, and use it to analyze data containing CGB from cosmic strings to study the detection threshold of cosmic string scenario for LISA.

The organization of the paper is as follows: in section \ref{sec:level2}, the background of cosmic string scenario is quickly reviewed, including the gravitational radiation from loops and stochastic gravitational-wave background from cosmic strings.  
In section \ref{sec:level4} we introduce the algorithm for model selection.  In section \ref{sec:level5} we utilize the model-selection algorithm to study the detection threshold of cosmic string scenario for LISA, and the analysis results will be presented.

\section{\label{sec:level2}Super-heavy Cosmic Strings}

In this section we will quickly summarize previous works, mainly by T. Vachaspati and A. Vilenkin, on cosmic string \cite{CosmicString,CosmicString3}.  
The expression for stochastic gravitational-wave background from cosmic strings used in this paper will be given.  

\subsection{Background Review}

Cosmic strings are topological defects which produced during 
a symmetry breaking phase transition in the very early universe.  
Although cosmic string scenario has been disfavored by CMBR data, 
it could be a significant source of gravitational-wave background. 
All strings are in the form of loops.  In a Hubble volume, there are a number of small loops, while there are few loops 
larger than a Hubble volume so that only a section of loops contained in the volume, which is called infinite strings.  
An exceptional feather of cosmic strings is that the number of infinite strings roughly remains constant during the evolution of the universe.  
The string networks evolve as the universe expands
These loops are characterized by mass density $\mu$.  In particular, the tension of loops are equal to $\mu$, 
which is much larger than the tension of daily-use strings.  Under such strong tension, the loops oscillate in a relativistic 
speed, which leads violent variations of mass distributions with respect to time, producing gravitational waves with typical frequency $f\sim L^{-1}$ where $L$ denotes as their lengths.  
The cosmic string networks develop with the expansion of the universe.  As the consequence, the number density of 
infinite strings roughly keep the same in a Hubble volume.  
Small loops would disappear due to releasing energy by gravitational waves.  
However, more loops would be produced from re-connection mechanism to replace the absent loops.  
T. Vachaspati and A. Vilenkin have made a careful calculation of gravitational radiation from 
oscillating closed-strings \cite{CosmicString} where the contributions from each loop 
in a wide range of frequencies was considered.  

Subsequently, it has been suggested that in brane inflation theory the fundamental (F-) and D-string networks could be produced during the evolution of universe.  The first new character of the superstrings is that their typical size may different from that of ordinary strings.  Secondly, the reconnection probability of intersecting superstrings may be much smaller than 1, while the ordinary strings always reconnect (p=1).  The two concerns lead to the reexamination of gravitational radiation from these cosmic superstrings \cite{CosmicString3}.  

\subsection{Gravitational Radiation from Loops}

Suppose the trajectory of the string is $\bold{x}(\sigma,t)$, $\sigma$ is a parameter along 
the string.  The equation of motion for a string \cite{CosmicString} is
\begin{equation}
\left\{ \begin{array}{r@{\quad \quad}l}
\dot{\mathbf{x}}-\mathbf{x}''=0\\
\dot{\mathbf{x}}\cdot\mathbf{x}'=0\\
\dot{\mathbf{x}}^{2}+\mathbf{x}'^{2}=1.
\end{array} \right.
\end{equation}
Dots and primes stand for derivatives with respect to $t$ and $\sigma$.  
A simple solution with only two frequencies is
\begin{eqnarray}
\mathbf{x}& = &\frac{L}{4\pi}\{\hat{e}_{1}[(1-\alpha)\sin\sigma_{-}+\frac{1}{3}\alpha\sin
3\sigma_{-}+\sin\sigma_{+}]\nonumber\\
&   & -\hat{e}_{2}[(1-\alpha)\cos\sigma_{-}+\frac{1}{3}\alpha\cos3\sigma_{-}+
\cos\phi\cos\sigma_{+}]\nonumber\\
& &-\hat{e}_{3}[2\sqrt{\alpha(1-\alpha)}\cos\sigma_{-}+\sin\phi\cos\sigma_{+}]\},
\end{eqnarray}
where plus and minus represent the left and right-moving modes of oscillations on 
loops.  This gives a fair representation of closed strings produced during the early universe.  
As high-frequency modes are decayed during the expansion of universe, the cutting 
off higher-order terms in Fourier series is a fair approximation.  
The distribution of gravitational radiation generated by a closed string is
\begin{equation}
\frac{dP_{n}}{d\Omega}=\frac{G\omega^{2}_{n}}{\pi}\left(T^{*}_{\mu\nu}
(\omega_{n},k)T^{\mu\nu}(\omega_{n},k)-\frac{1}{2}|T^{\mu}_{\mu}(\omega_{n},k)|
^{2}\right),
\end{equation}
where $\omega_{n}=4\pi n/L$ is the angular frequency, and $T_{\mu\nu}(\omega_{n},k)$ 
is the Fourier transform of the stress-energy tensor of a string
\begin{equation}
T_{\mu\nu}(\mathbf{x},t)=\mu\int{d\sigma\left(\dot{\mathbf{x}}^{\mu}
\dot{\mathbf{x}}^{\nu}-\mathbf{x}'^{\mu}\mathbf{x}'^{\nu}\right)\delta^{(3)}
\left(\mathbf{x}-\mathbf{x}(\sigma,t)\right)},
\end{equation}
$\mu$ is mass per unit length.  The total power of the gravitational radiation produced by the 
string is
\begin{equation}
\frac{dE}{dt}=\sum_{n}P_{n},
\end{equation}
where 
\begin{equation}
P_{n}=\int{\left(\frac{dP_{n}}{d\Omega}\right)d\Omega}.
\end{equation}  
It can also be expressed by $\gamma G\mu^{2}$, where $\gamma\sim 50$ is derived 
from numerical computation.

\subsection{Stochastic Gravitational-wave Background from Cosmic Strings}

The stochastic gravitational background was produced by loops with different shapes and sizes 
around us.  The formula of total gravitational radiation by a loop has been derived.  The 
energy density of the gravitational background can be calculated by integrating the product of 
the total power from a loop and the number density over entire history of universe, as long as 
the number density of loops is suggested.
    
Loops only emit gravitational waves at specific frequencies $f_{n}=2n/L(t)$ relating with 
their length.  The $L(t)=L_{n}-\gamma G\mu t$ is their length, where $\gamma G\mu$ 
represents their contraction caused by the gravitational radiation.  Their 
formation rate $dn/dt$ at time $t$ and their initial length $L_{n}$ are proportional to 
$\beta t^{-4}$ and $\alpha t$, respectively, where $\alpha$ and $\beta$ are numerical 
coefficients.  By the relation mentioned above, it is shown that the number density of strings 
corresponding to the gravitational waves' frequency $f$ observed presently is
\begin{equation}
dn_{n}(t)=\alpha^{3}\beta\left(\frac{\alpha(L_{n}/\alpha)}{a(t)}\right)^{3}
\frac{dL_{n}}{L^{4}_{n}},
\end{equation}
where the scale factor $a(t)$ represents the expansion of universe.  The present energy density 
of the radiation with frequency $f$ is
\begin{equation}
d\rho_{g}=\frac{dE}{dt}\int^{t_{0}}{\left(\frac{a(t)}{a(t_{0})}\right)^{4}
dt\,dn_{n}}.
\end{equation}
The lower bound of the integration is the time $t\sim L_{n}/\alpha$ when the loop was produced. 

Wyman, Pogosian, and Wasserman found the constraints from WMAP and SPSS data 
on the fraction of cosmological fluctuations by cosmic strings \cite{StringBound1}.  
Their results suggested that cosmic strings can account for up to 7 (14\%) of the total 
power of the microwave anisotropy at 68 (95\%) confidence level.  The corresponding 
bound on $G\mu$ is $3.4\ (5)\times 10^{-7}$ at 68 (95\%) confidence level.  
M. V. Sazhin, O. S. Khovanskaya, M. Capaccioli, G. Longo, J. M. Alcala, R.
Silvotti, and M. V. Pavlov found two nearly identical galaxies are having 
angular separation of 1.9 arc sec, which suggests $G\mu\sim 4\times 10^{-7}$ 
if the gravitational lensing is caused by a cosmic string \cite{StringBound2}.  
Janet et al. \cite{StringBound3} developed a technique to detect a stochastic 
gravitational-wave background by finding correlations between pulsar observations.  
Based on this work, a method to place an upper bound on the power of a specific stochastic 
gravitational-wave background (according to the corresponding model function) 
by using observations of multiple pulsars was presented \cite{StringBound4}.  
Eight year of millisecond pulsar timing observations 
has provided a limit on the critical density of gravitational wave 
from cosmic strings: $\Omega_{GW}h^{2}\le 1.9\times 10^{-8}$.


\begin{figure}
\includegraphics[scale=0.31]{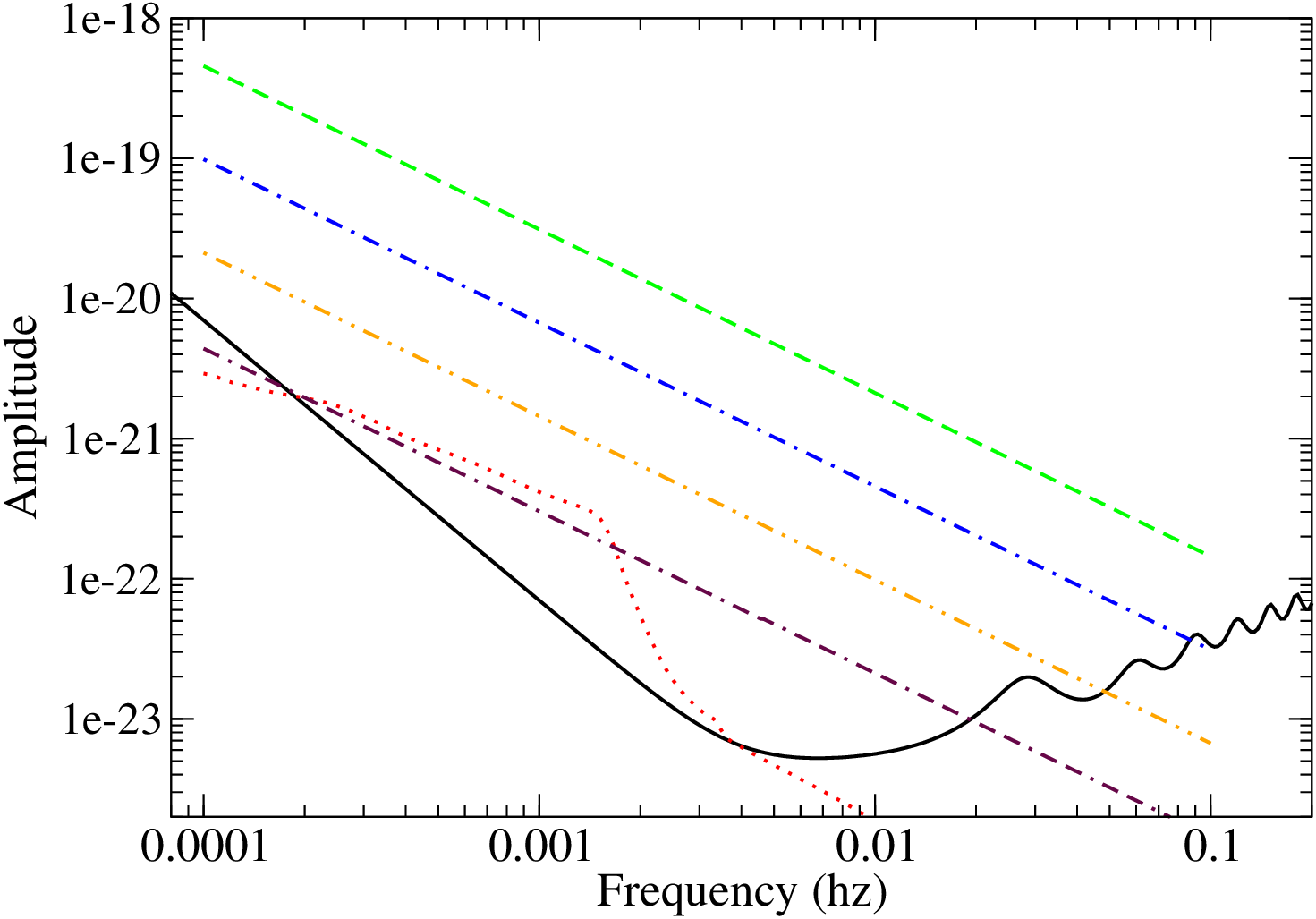}
\caption[Gravitational-wave amplitude by cosmic strings with various values of $G\mu$ 
               against frequency]{
 The black (solid) curve shows LISA sensitivity.  The red (dot) curve represents the confusion 
 noise background.  The green (dash), blue (dash-dot), yellow (double-dot dash), and 
 purple (double-dash dot) lines 
 illustrate the the amplitude for $G\mu$ of $10^{-7}$, $10^{-9}$, $10^{-11}$, and $10^{-13}$, respectively, 
 in super-heavy cosmic strings scenario.  
 $\epsilon$ and p are kept as 1.
 The figure suggests that LISA may be able to detect the gravitational 
 waves in the scenario as $G\mu$ is down to $10^{-13}$.}
\label{fig:Gmu}
\end{figure}

The stochastic gravitational-wave background from cosmic strings is composed of the waves 
emitted by the string loops across the sky.  In \cite{CosmicString3} the stochastic 
background is referred to `confusion noise background' which is defined by
\begin{equation}
h_{confusion}^{2}(f) = \int \frac{dz}{z}n(f,z)h^{2}(f,z)\Theta(n(f,z)-1).
\label{eq:h-confusion-string}
\end{equation}
$n(f,z)$, the number of the string loops in the universe at frequency $f$ and redshift 
$z$, is given by
\begin{equation}
n(f,z) = 100 \frac{c\epsilon}{p}(ft_{0})^{-5/3}\alpha^{-8/3}\varphi_{n}(z)C(z), 
\label{eq:n-calculation}
\end{equation}
and the dimensionless amplitude of gravitational wave produced by a single loop is
\begin{equation}
h(f,z)=G\mu\alpha^{2/3}(ft_{0})^{-1/3}\varphi_{h}(z)\Theta(1-\theta_{m}
[\alpha,f,z])
\label{eq:h-calculation}
\end{equation}
where $\alpha$ parametrises the typical size of a loop, $t_{0}$ is present cosmological time 
$\approx 4.42\times 10^{17}$ second, and $z_{eq} = 10^{3.94}$ is the redshift of equal 
matter and radiation densities.  
Here
\begin{equation}
\varphi_{n}(z) = z^{3}(1+z)^{-7/6}(1+\frac{z}{z_{eq}})^{11/6},
\end{equation}
\begin{equation}
\varphi_{h}(z) = z^{-1}(1+z)^{-1/3}(1+\frac{z}{z_{eq}})^{-1/3},
\end{equation}
and
\begin{equation}
C(z) = 1+ \frac{9z}{z+z_{eq}}
\end{equation}
is an interpolation between matter-dominant era and radiation-dominant era.  
Bear in mind that the step function $\Theta(1-\theta_{m}[\alpha, f, z])$ serves as a 
cut-off to the Fourier component whose mode number is less than one in which 
$\theta_{m}[\alpha,f,z]$ is given by 
\begin{equation}
\theta_{m}(\alpha,f,z) = (\alpha ft_{0})^{-1/3}(1+z)^{1/6}
(1+\frac{z}{z_{eq}})^{1/6}.
\end{equation}
\begin{figure}
\includegraphics[scale=0.31]{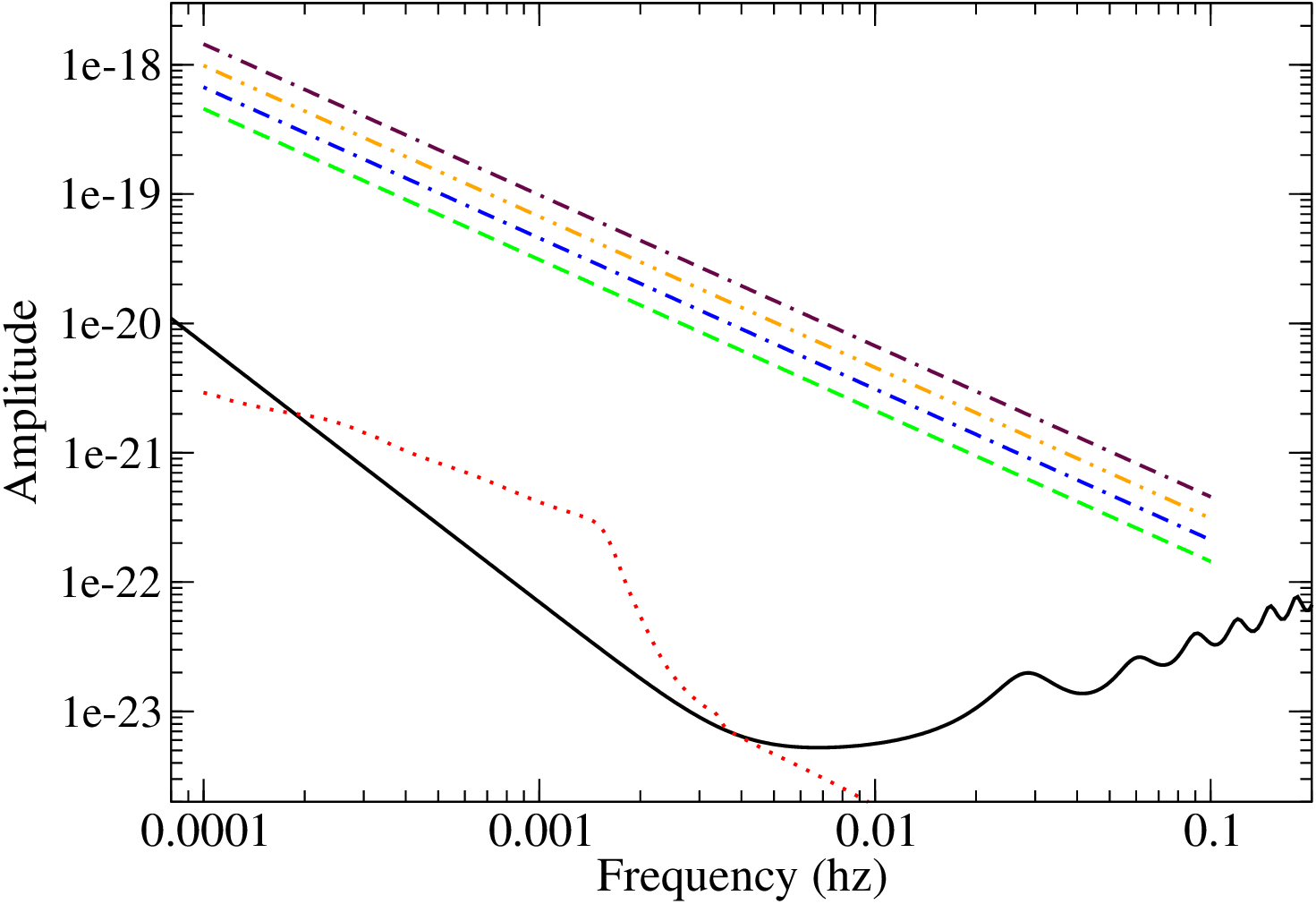}
\caption[Gravitational-wave amplitude by cosmic strings with various values of $\epsilon$ 
               against frequency]{
 The black (solid) curve shows LISA sensitivity.  The red (dot) curve represents the confusion 
 noise background.  The green (dash), blue (dash-dot), yellow (double-dot dash), and 
 purple (double-dash dot) lines 
 illustrate the the amplitude for $\epsilon$ of 1, 0.1, 0.01, and 0.001, respectively, 
 in super-heavy cosmic strings scenario.  
 $G\mu$ is kept as $10^{-7}$, and p is fixed as 1.  
 The figure displays that $\epsilon$ does not affect on gravitational-wave 
 amplitude very much.}
\label{fig:epsilon}
\end{figure}

\begin{figure}
\includegraphics[scale=0.31]{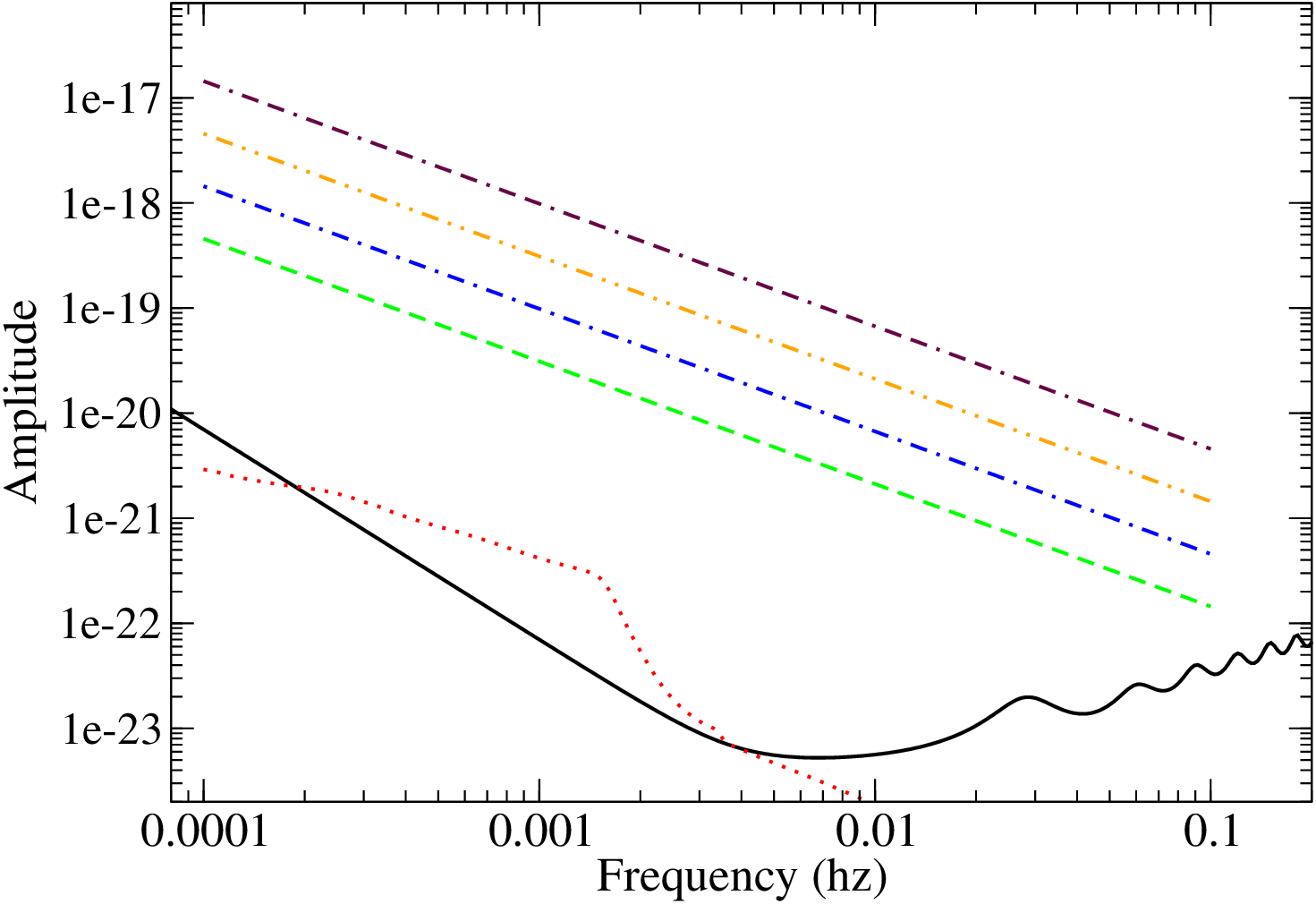}
\caption[Gravitational-wave amplitude by cosmic strings with various values of 
               reconnection probability p against frequency]{
 The black (solid) curve shows LISA sensitivity.  The red (dot) curve represents the confusion 
 noise background.  The green (dash), blue (dash-dot), yellow (double-dot dash), and 
 purple (double-dash dot) lines 
 illustrate the the amplitude for reconnection probability p of 1, 0.1, 0.01, and 0.001, respectively, 
 in super-heavy cosmic strings scenario.  
 $G\mu$ is kept as $10^{-7}$, and $\epsilon$ is fixed as 1.  
 }
\label{fig:p}
\end{figure}

The step function $\Theta(n(f,z)-1)$ constrains the integration over $z$ in the case of 
multiple sources, which are regarded as the origin of confusion.  
However, this consideration is not appropriate for LISA.  
This is because LISA is designed to detect continuous sources.  
In order to reconstruct short-duration events such as bursts by triangulation, 
the minimum number of independent detectors is 3.  
For the reason, we adopt root-mean-square definition 
\begin{equation}
P_{string}(f) = h^{2}_{rms}(f) = \int \frac{dz}{z}n(f,z)h^{2}(f,z),
\label{eq:h-rms-string}
\end{equation} 
averaged over all events, to calculate the power spectral density of the stochastic background.  
Substituting Eq.\ \eqref{eq:n-calculation} and \eqref{eq:h-calculation} into 
Eq.\ \eqref{eq:h-rms-string}, we have
\begin{eqnarray}
P_{string}(f) &=& \frac{100}{50^{4/3}}\frac{c}{p}\epsilon^{-1/3}(G\mu)^{2/3}
t^{-7/3}_{0}\nonumber\\
&\times&\int\frac{(1+z/z_{eq})^{7/6}}{(1+z)^{11/6}}dz\ f^{-7/3}, 
\label{eq:hrms}
\end{eqnarray}
where $t_{0}=4.4\times 10^{17}$ and $c=1$.  The integration can be 
worked out numerically, and its value is 0.21.  
Then the Eq.\ \eqref{eq:hrms} can be reduced to 
\begin{equation}
P_{string}(f) = 3.65\times10^{-42}\frac{(G\mu)^{2/3}}{p\epsilon^{1/3}}\ f^{-7/3}.
\label{eq:hrms-analysis}
\end{equation}
Figure \ref{fig:Gmu}, \ref{fig:epsilon}, and \ref{fig:p} illustrate the influences of the 
three parameters, $G\mu$, $\alpha$, and $p$, on the strain amplitude h.

\section{\label{sec:level4}Algorithm for Model Selection}
\label{sec:model-selection}

Suppose that there are two models $M_{1}$ and $M_{2}$ which are the possible 
explanations for a data set.  
The ratio $P(M_{1}|D,I)/P(M_{2}|D,I)$, namely the Bayes factor $B_{12}$, is regarded as 
the relative preference of $M_{1}$ over $M_{2}$.  Determining the better explanation is 
identical to find the Bayes factor.

Suppose that the two models $M_{1}$ and $M_{2}$ involve $r$ parameters 
$p_{1},\cdots,p_{r}$ and $s$ parameters $q_{1},\cdots,q_{s}$, respectively where $s>r$.  
With the help of Baye's theorem, the relative probability $P(M_{1}|D,I)/P(M_{2}|D,I)$ 
can be decomposed to the product of relative likelihood function and prior
\begin{equation}
B_{12}=\frac{P(M_{1}|\ D,I)}{P(M_{2}|\ D,I)} = \frac{P(D|\ M_{1},I)}{P(D|\ M_{1},I)}
\times\frac{P(M_{2}|\ I)}{P(M_{2}|\ I)}.
\end{equation}
The $P(M_{1}|I)$ is the prior of model $M_{1}$ being true for given background 
information.  Here the evidence is cancelled out because the identical data set is used.  
Since we do not prefer any specific model before we analyse data, 
the prior ratio $P(M_{1}|I)/P(M_{2}|I)$ can be set to 1.  
$P(D|\ M_{1},I)$ is marginal likelihood function
\begin{equation}
P(D|\ M_{1},I) = \int P(D,p_{1},\cdots,p_{r}|M_{1},I)\ dp_{1}\cdots dp_{r}.
\label{eq:model-selection-1}
\end{equation}
With Baye's theorem, $P(D,p_{1},\cdots,p_{r}|M_{1},I)$ can be further separated to 
the product of likelihood function $P(D|\ p_{1},\cdots,p_{r},M_{1},I)$ and prior 
$P(p_{1},\cdots,p_{r}|\ M_{1},I)$.  
Hence, Eq. \eqref{eq:model-selection-1} can be expressed as 
\begin{eqnarray}
P(D|\ M_{1},I) &=& \int P(D|\ p_{1},\cdots,p_{r},M_{1},I)\nonumber\\
&\times& P(p_{1},\cdots,p_{r}|\ M_{1},I)\ dp_{1}\cdots dp_{r}.\
\label{eq:marginal-likelihood}
\end{eqnarray}

Presumably, any particular set of values for $(p_{1},\cdots,p_{r})$ is not favoured before 
analysis.  
Thus, the prior can be assigned by a very wide multiple Gaussian distribution
\begin{eqnarray}
P(p_{1},\cdots,p_{r}|\ M_{1},I) &=& \frac{1}{\sqrt{2\pi}\delta_{1}}\exp\Big\{
-\frac{p^{2}_{1}}{2\delta^{2}_{1}}\Big\}\times\cdots\nonumber\\
&\times&\frac{1}{\sqrt{2\pi}\delta_{r}}\exp\Big\{-\frac{p^{2}_{r}}{2\delta^{2}_{r}}\Big\}.
\label{eq:prior}
\end{eqnarray} 

The likelihood function $P(D|\ p_{1},\cdots,p_{r},M_{1},I)$ can be approximated 
around its best estimate $(\bar{p_{1}},\cdots,\bar{p_{r}})$
\begin{eqnarray}
P(D&|&\ p_{1},\cdots,p_{r},M_{r},I) \approx 
P(D|\ \bar{p_{1}},\cdots,\bar{p_{r}},M_{1},I)\nonumber\\
&\times&\exp\left\{-\frac{1}{4}(\mathbf{p}-\bar{\mathbf{p}})^{T}
\mathbf{\sigma_{1}}^{-1}(\mathbf{p}-\bar{\mathbf{p}})\right\}
\label{eq:posterior}
\end{eqnarray}
where $\mathbf{\sigma_{1}}$ is the covariant matrix, $\mathbf{p}=(p_{1},\cdots,p_{r})$, and 
$\bar{\mathbf{p}}=(\bar{p_{1}},\cdots,\bar{p_{r}})$.  
Since the uncertainty of the prior is very board, the centre of prior can be shifted to the 
best estimate 
\begin{eqnarray}
P(p_{1},\cdots,p_{r}&|&\ M_{1},I) \approx \frac{1}{\sqrt{2\pi}\delta_{1}}
\times\cdots\times\frac{1}{\sqrt{2\pi}\delta_{r}}\nonumber\\
&\times&\exp\Big\{-\frac{1}{2}(\mathbf{p}-\bar{\mathbf{p}})^{T}
\mathbf{\Delta}^{-1}(\mathbf{p}-\bar{\mathbf{p}})
\Big\}.
\end{eqnarray}
Here the $\mathbf{\Delta}^{-1}$ is a diagonal matrix where the $i$-th term is $\delta^{2}_{i}$.  
Multiplying the prior and the likelihood function, we can get 
\begin{widetext}
\begin{eqnarray}
P(p_{1},\cdots,p_{r}|\ M_{1},I)\times P(D|\ p_{1},\cdots,p_{r},M_{1},I) &\approx&
P(D|\ \bar{p_{1}},\cdots,\bar{p_{r}},M_{1},I)\times \frac{1}{\sqrt{2\pi}
\delta_{1}}\times\cdots\times\frac{1}{\sqrt{2\pi}\delta_{r}}\nonumber\\
&\times&\exp\Big\{-\frac{1}{4}(\mathbf{p}-\bar{\mathbf{p}})^{T}
(\sigma^{-1}_{1}+2\mathbf{\Delta}^{-1})(\mathbf{p}-\bar{\mathbf{p}})\Big\}.
\label{eq:prior-likelihood}
\end{eqnarray}
\end{widetext}
$\Delta^{-1}$ only affects the diagonal term of $\sigma^{-1}$, and the $i$-th diagonal term 
of the matrix $\sigma^{-1}_{1}+2\Delta^{-1}$ is 
\begin{equation}
\frac{1}{\sigma^{2}_{1\ ii}}+\frac{2}{\delta^{2}_{ii}} =
\frac{\delta^{2}_{ii}+2\sigma^{2}_{1\ ii}}{\sigma^{2}_{1\ ii}\delta^{2}_{ii}}\approx
\frac{1}{\sigma^{2}_{1\ ii}}
\end{equation}
since $\delta_{ii}\gg \sigma_{1\ ii}$.  Hence, Eq.\ \eqref{eq:prior-likelihood} can be 
approximated to 
\begin{widetext}
\begin{eqnarray}
P(p_{1},\cdots,p_{r}|\ M_{1},I)\times P(D|\ p_{1},\cdots,p_{r},M_{1},I) &\approx&
P(D|\ \bar{p_{1}},\cdots,\bar{p_{r}},M_{1},I)\times \frac{1}{\sqrt{2\pi}
\delta_{1}}\times\cdots\times\frac{1}{\sqrt{2\pi}\delta_{r}}\nonumber\\
&\times&\exp\Big\{-\frac{1}{4}(\mathbf{p}-\bar{\mathbf{p}})^{T}
\sigma^{-1}_{1}(\mathbf{p}-\bar{\mathbf{p}})\Big\}.
\label{eq:prior-likelihood-1}
\end{eqnarray}
\end{widetext}
To evaluate the integral in Eq. \eqref{eq:marginal-likelihood} we need to use the integral of 
$r$-dimensional multivariate Gaussian:
\begin{eqnarray}
&&\int\cdots\int \exp\Big\{-\frac{1}{4}(\mathbf{p}-\bar{\mathbf{p}})^{T}
\sigma^{-1}_{1}(\mathbf{p}-\bar{\mathbf{p}})\Big\}\ dp_{1}\cdots dp_{r}\nonumber\\
&&=(4\pi)^{r/2}\sqrt{Det(\sigma_{1})}.
\label{eq:int-multi-Gaussian}
\end{eqnarray}
With Eq.\ \eqref{eq:int-multi-Gaussian}, substituting Eq. \eqref{eq:prior-likelihood-1} 
into Eq. \eqref{eq:marginal-likelihood} we can obtain
\begin{equation}
P(D|\ M_{1},I) \approx \frac{2^{r/2}\sqrt{Det(\sigma_{1})}}
{\delta_{1}\cdots\delta_{r}}\ P(D|\bar{p_{1}},\cdots,\bar{p_{r}},M_{1},I).
\end{equation}
In principle, the values of $\delta_{1},\cdots,\delta_{r}$ are considered as infinite to 
satisfy the condition of the prior being very board.  In practice, there is a way to assign 
fairly large values for the uncertainties of prior.  For instance, the uncertainties can be set by 
the range of the parameters, which can be known either from the experiment design or
the results of previous experiment, or guessed from theoretical calculation.  

By repeating the same steps, we can have the expression for $P(D|\ M_{2},I)$, and then we 
can acquire the relative probability of $M_{2}$ and $M_{1}$
\begin{eqnarray}
\frac{P(D|\ M_{2},I)}{P(D|\ M_{1},I)} &\approx& \sqrt{2^{s-r}}\ 
\frac{\delta_{1}\cdots\delta_{r}\sqrt{Det(\sigma_{2})}}{\delta^{'}_{1}\cdots\delta^{'}_{s}
\sqrt{Det(\sigma_{1})}}\nonumber\\
&\times&
\frac{P(D|\bar{p_{1}}^{'},\cdots,\bar{p_{s}}^{'},M_{2},I)}
{P(D|\bar{p_{1}},\cdots,\bar{p_{r}},M_{1},I)},
\label{eq:model-comparison}
\end{eqnarray}
where the symbol prime refers to the objects related to $M_{2}$ model, and 
$\sigma_{1}$ and $\sigma_{2}$ are the covariant matrixes given by fitting with 
$M_{1}$ and $M_{2}$ model respectively.  
Since the noise is independent and additive, 
the likelihood function $P(D|\bar{p_{1}}^{'},\cdots,\bar{p_{s}}^{'},M_{2},I)$ and 
$P(D|\bar{p_{1}},\cdots,\bar{p_{r}},M_{1},I)$ can be decomposed into a series 
product of Gaussian distributions.
%
 %
Table \ref{Bayes-factor} lists the evidence for 
model $M_{1}$ against the Bayes factor.
\begin{table}
\begin{center}
\caption{Bayes factor confidence level cited from \cite{Model-selection}}
\begin{tabular}{lc}
 \hline\hline $B_{12}$ & Evidence for model $M_{1}$\\ \hline      
 \raisebox{0pt}[2.5ex]{}$<1$  & \vspace{0.1cm}Negative \\
 \raisebox{0pt}[2.5ex]{}$1\sim 3$ & \vspace{0.1cm}Not worth more than a bare mention \\
 \raisebox{0pt}[2.5ex]{}$3\sim 12$ & \vspace{0.1cm}Positive \\
 \raisebox{0pt}[2.5ex]{}$12\sim 150$ & \vspace{0.1cm}Strong \\
 \raisebox{0pt}[2.5ex]{}$> 150$ & \vspace{0.1cm}Very Strong \\
 \hline \hline
\end{tabular}
\end{center}
\label{Bayes-factor}
\end{table}

If the noise $\sigma$ is unknown, we can integrate the series product of 
Gaussian distributions over $\sigma$ to obtain the marginalised likelihood 
function, which is 
\begin{equation}
P(D|\bar{p_{1}},\cdots,\bar{p_{r}},M_{1},I) = \frac{(N-1)!}{N^{N}}
\Big(\frac{F(\bar{p_{1}},\cdots,\bar{p_{r}})}{N}\Big)^{-(N-1/2)}
\end{equation}
where $F(\bar{p_{1}},\cdots,\bar{p_{r}})$ is defined as 
$\sum^{N}_{i=1}(p^{d}_{i}-p^{f}_{i}(\bar{p_{1}},\cdots,\bar{p_{r}}))$.  
The other marginalised likelihood function 
$P(D|\bar{p_{1}}^{'},\cdots,\bar{p_{s}}^{'},M_{2},I)$ is given by the same step.  
With the two marginalised likelihood functions, we have the likelihood ratio 
\begin{equation}
\frac{P(D|\bar{p_{1}}^{'},\cdots,\bar{p_{s}}^{'},M_{2},I)}
{P(D|\bar{p_{1}},\cdots,\bar{p_{r}},M_{1},I)}
=\left(\frac{F(\bar{p_{1}},\cdots,\bar{p_{r}})}
{F(\bar{p_{1}}^{'},\cdots,\bar{p_{s}}^{'})}\right)^{N-1/2}
\label{eq:marginalised-likelihood-ratio}
\end{equation}
in the case that the noise is unknown.  
If the correlations among parameters are small, $\sqrt{Det(\sigma_{1})}$ can be further 
approximated to successive product of uncertainties 
$\sigma_{p_{1}}\cdots\sigma_{p_{r}}$.  
We define the small as the magnitude of correlation being smaller than 0.2, 
in which case the discrepancy between the approximation and the true value of determinant 
is about $5\%$.  
Expanding $\sqrt{Det(\sigma_{2})}$ as 
$\sigma^{'}_{p_{1}}\times\cdots\times\sigma^{'}_{p_{s}}$, 
the first term on the right hand side of Eq. \eqref{eq:model-comparison} can be simplified as 
\begin{equation}
\sqrt{2^{s-r}}\ \frac{\delta_{1}\cdots\delta_{r}\sqrt{Det(\sigma_{2})}}
{\delta^{'}_{1}\cdots\delta^{'}_{s}\sqrt{Det(\sigma_{1})}}
=\sqrt{2^{s-r}}\ \frac{\delta_{1}}{\sigma_{p_{1}}}\cdots
\frac{\delta_{r}}{\sigma_{p_{r}}}\cdots
\frac{\sigma^{'}_{p_{1}}}{\delta^{'}_{1}}\cdots
\frac{\sigma^{'}_{p_{s}}}{\delta^{'}_{s}}.
\label{eq:Ockham}
\end{equation}
This term is the so-called \emph{Ockham factor}.  To understand this factor, we suppose that 
the model $M_{2}$ is built upon $M_{1}$ with an extra parameter $p_{r+1}$.  
The uncertainties of parameters estimated by different models are roughly the same, 
so the right hand side of Eq.\ \eqref{eq:Ockham} 
can be approximated by 
\begin{equation}
\sqrt{2^{s-r}}\ \frac{\delta_{1}\cdots\delta_{r}\sqrt{Det(\sigma_{2})}}
{\delta^{'}_{1}\cdots\delta^{'}_{s}\sqrt{Det(\sigma_{1})}}
\approx \sqrt{2}\ \frac{\sigma^{'}_{p_{r+1}}}{\delta^{'}_{r+1}}.
\end{equation}
The relative probability of $M_{2}$ and $M_{1}$ can be simplified as 
\begin{equation}
\frac{P(D|\ M_{2},I)}{P(D|\ M_{1},I)} \approx \sqrt{2}\ 
\frac{\sigma^{'}_{p_{r+1}}}{\delta^{'}_{r+1}}\times
\frac{P(D|\bar{p}^{'}_{1},\cdots,\bar{p}^{'}_{r+1},M_{2},I)}
{P(D|\bar{p_{1}},\cdots,\bar{p_{r}},M_{1},I)}.
\label{eq:model-comparison-1}
\end{equation}
The second term in Eq. \eqref{eq:model-comparison-1} is the likelihood ratio of two 
models.  
Since a more complex model will yield a better fit than a simpler model, 
the likelihood ratio in Eq.\ \eqref{eq:model-comparison-1} is always larger than 1.  
On the other hand, it is found that the Ockham factor is much smaller than 1 as long as 
the quality of data is not poor.  
(poor-data limit can be expressed as $\sigma^{'}_{r+1} \gg \delta^{'}_{r+1}$.)
It plays a role of `penalty' for using an extra parameter to fit data.  
As we can see, if the more complex model cannot produce a much better fitting, giving a 
high likelihood ratio to overcome the penalty, the simpler model will be suggested.  
Although a more complex model always fits data better, 
a simpler one might be preferred if the discrepancy of its descriptions to data 
is not much worse than the more complex ones.

To make a judgement on the number of signals in data, we can use the simplest model 
$M_{1}$ which assumes that there is no signal contained in data, and the model $M_{2}$ 
which assumes that there is one signal in the data to analyse the data, respectively.  
Then we compare their relative probability $P(M_{2}|D,I)/P(M_{1}|D,I)$.  If the ratio 
is smaller than 1, it indicates that the data are just purely noisy; otherwise, it is suggested 
that there may be one or more signals out there.  We further move to the more complex model 
$M_{3}$ and compute the ratio $P(M_{3}|D,I)/P(M_{2}|D,I)$.  If the ratio is smaller than 1, 
it is suggested that there is one signal in the data; otherwise, we move forward to the more 
complex model $M_{4}$ and follow the same steps to compute the ratio and so on 
until it is smaller than 1.  The likeliest model would advise us the best description of the data.

In the end, we would like to discuss the uncertainty of prior $\delta^{'}_{r+1}$ in 
Eq.\ \eqref{eq:model-comparison-1} deeper.  A concern is that the uncertainty seems to be 
free to choose.  At least, various approaches would propose different values of the uncertainty.  
Therefore, $M_{1}$ model may be inclined with the choices of some values, 
but $M_{2}$ model may be indicated with the others.  
In other words, the model selection is worried to be subjective.  
To clear this unease, we should keep in mind that choosing a bigger value of the uncertainty 
just represents the ground is more conservative.  The bigger uncertainty might lead to reject the 
signal which is indicated by a model with a smaller uncertainty.  However, it is just shown 
that the signal is not strong enough to be accepted by the conservative point of view.  
The relative probability $P(M_{2}|D,I)/P(M_{1}|D,I)$ just can be considered as the relative 
degree of belief between the two models.  
After all, we choose to regard probability as the degree of belief rather than a long term 
occurrence rate as we use Bayesian statistics in the first place.
%
%

\section{\label{sec:level5}Data Analysis}
\begin{figure}
\includegraphics[scale=0.31]{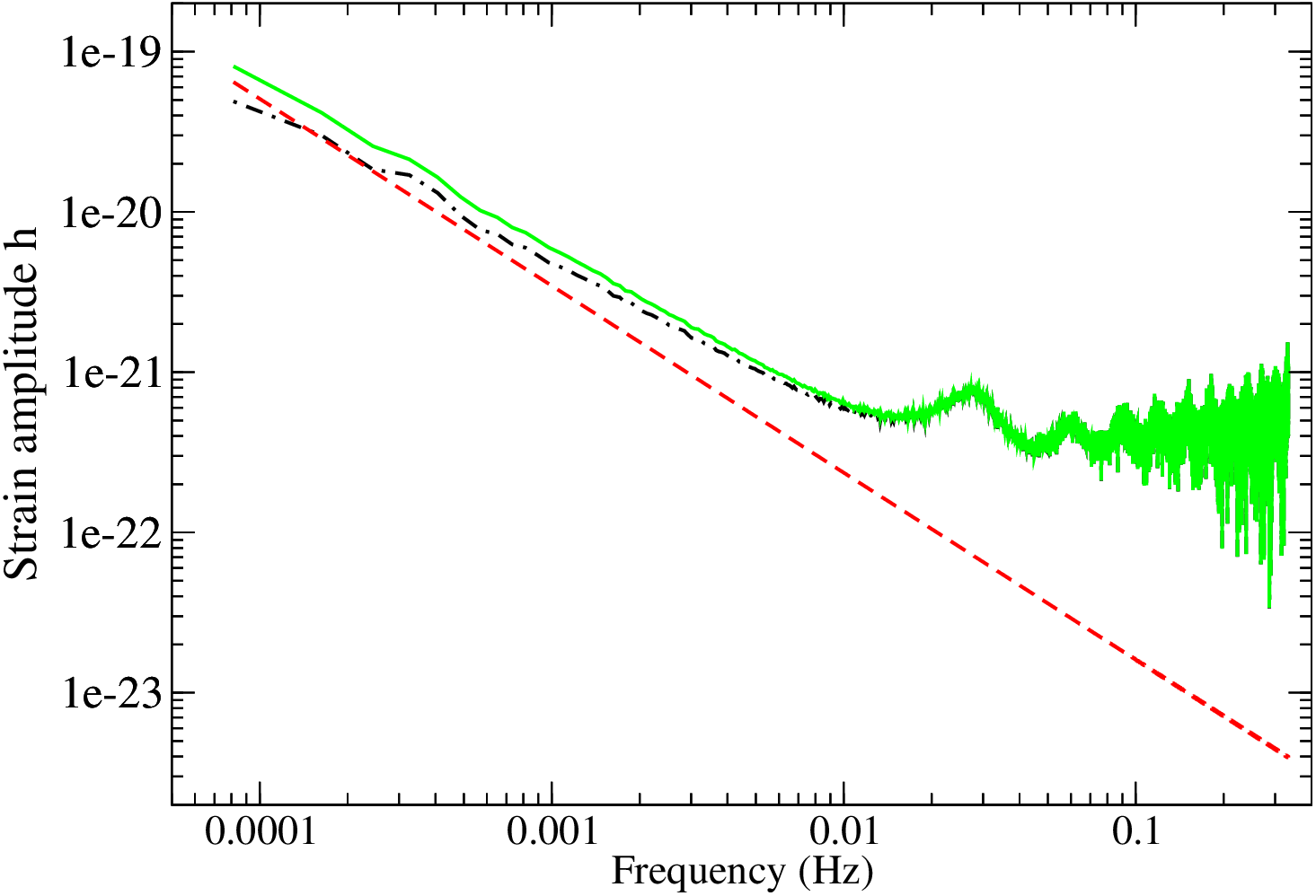}
\caption[Comparison of gravitational-wave amplitude from cosmic strings between by
               simulation and by Eq.\ \eqref{eq:h-confusion-string}]{
 On the top plot the black solid curve is the simulation of random acceleration and shot noise 
 on the `sky plane'.  
 The red dash line is the cosmic gravitatinal-wave background produced by cosmic strings. 
 On the bottom plot the black solid curve is the full spectrum including both the background, 
 the random acceleration, and the instrumental noise plotted on the `sky plane'.
}
\label{fig:string-simulation}
\end{figure}

To apply the model selection algorithm to determine the existence of the gravitational-wave 
background generated by the cosmic strings in a spectrum, we need two models 
to describe the spectrum in which the background is present or absent, respectively.  
Models were constructed in the hierarchical sense, which means that all models are built on a primary one.  
In the case of the background from cosmic strings being absent, the spectrum only contains the power from 
the drift trend due to random acceleration.  
Because the trend definitely exist, it is chosen as our primary model.  
The model, namely $M_{0}$, can be expressed by Eq.\ \eqref{eq:M-0}:
\begin{equation}
M_{0}(f_{n})=\frac{N\Delta^{4}a^{2}}{4L^{2}}\Big(\frac{1}{\sin^{4}
\frac{\pi n}{N}}+\frac{N^{2}-1}{3\sin^{2}\frac{\pi n}{N}}\Big)
\label{eq:M-0}
\end{equation}
where N, the total number of data, is set as 8192 in our research, n is the index.  
$\Delta$, the sampling rate, is  set as 1.5 sec, and $f_{n}=n/N\Delta$ is the frequency.  
The total observation time $N\Delta$ is 12288 sec.  
The data from this setup can cover the LISA frequency band 0.1 Hz to 0.1 mHz.  
L is the arm-length of LISA, and a is the averaged acceleration need to be estimated.  

The complex model is established by adding a specific kind of signal to the simpler one.   
In the case of the background from cosmic strings being present, 
the spectrum is contributed from the drift trend due to random acceleration as well as the background.  
As for the background, the physical parameters $p$, $\epsilon$, and $G\mu$ in the 
Eq.\ \eqref{eq:h-rms-string} only change the strength of the spectrum 
but do not influence the spectrum shape.  
We cannot determine these three parameters uniquely through the strength of 
gravitational-wave background by the cosmic strings unless alternative information is acquired.  
Hence, we define a new parameter A in Eq.\ \eqref{eq:hrms} to describe their 
combined influences on amplitude as 
\begin{equation}
A \equiv 3.65\times10^{-42}\frac{(G\mu)^{2/3}}{p\epsilon^{1/3}}\ Hz^{7/3}.
\label{eq:A-value}
\end{equation}
Then the expression for analysing the spectrum is reduced to 
\begin{equation}
P_{string}(f_{n}) = A\ f^{-7/3}_{n}.
\label{eq:hrms-analysis}
\end{equation}
The model for the power spectra including both of random acceleration and 
cosmic string background, $M_{1}(f_{n})$, can be expressed by
\begin{equation}
M_{1}(f_{n}) = \frac{N\Delta^{4}a^{2}}{4L^{2}}\left[\frac{1}{\sin^{4}
\frac{\pi n}{N}}+\frac{N^{2}-1}{3\sin^{2}\frac{\pi n}{N}}\right]\frac{1}{R(f_{n})}
+A\ f^{-7/3}_{n}.
\label{eq:power-string+noise}
\end{equation}  

The spectrum produced by the cosmic strings was synthesised with LISA instrumental noise 
on the `sky plane' to produce full spectra containing both of signal and noise as follows:
\begin{equation}
P_{d}(f)=P_{string}(f)+\frac{P_{Ins}(f)}{R(f)}
\end{equation}
where $P_{Ins}(f)$ is the simulated power spectrum due to the LISA shot noise and the 
random acceleration, and $R(f)$ is the response function of LISA to stochastic backgrounds.  
The $P_{d}(f)$ is the spectra we used to test the algorithm.  An example of the 
simulated full spectrum is shown in the Figure \ref{fig:string-simulation}.

The data analysis starts from parameter estimation by the primary model $M_{0}$.  
Then the model $M_{1}$ is utilized to repeat the process.  
Their results are compared through eq. \eqref{eq:model-comparison-1}. 
If the model $M_{0}$ is preferred, it is indicated that there is no signal in the data set; 
otherwise, the background from cosmic strings may exist. 

\subsection{Upper Bound for the combined amplitude A}
In the model $M_{0}(f)$, the magnitude of random acceleration $a$ is the only parameter.  
As for $M_{1}(f)$, the model includes two parameters, the magnitude of random 
acceleration $a$ and the strength of the gravitational-wave background by cosmic strings $A$.  
To apply Eq.\ \eqref{eq:model-comparison} to compare the model $M_{0}(f)$ and 
$M_{1}(f)$, we have to know $\delta_{a}/\delta_{a}\delta_{A}=1/\delta_{A}$ 
where $\delta_{A}$ is regarded as the allowable range for the parameter $A$
, Max(A)-Min(A).  Since the minimal value of amplitude is zero, to determine $\delta A$, 
the rest thing we must know is the upper bound of $A$.  

The parameter $A$ is associated with the energy density of the gravitational-wave background 
through Eq. (4.7) in \cite{CosmicString3}:
\begin{equation}
\Omega_{g}(f)\sim \frac{3\pi^{2}}{2}(ft_{0})^{2}h^{2}(f)
=2.87\times10^{36}Af^{-1/3}.
\end{equation}
Therefore, giving an upper limit $U$ of $\Omega_{g}(f)$, the upper bound of $A$ can be 
obtained as 
\begin{equation}
A < 3.49\times10^{-37}Uf^{1/3} \ Hz^{7/3}.
\end{equation}
The upper limit $G\mu < 6.1\times 10^{-7}$ given by the CMB data 
\cite{StringBound1, StringBound2} yields the upper bound 
$A < 2.63\times 10^{-46}\ Hz^{7/3}$ through Eq.\ \eqref{eq:A-value}.  
The analysis of 8 year of millisecond pulsar-timing data provided 
$\Omega_{g}h^{2} < 6\times 10^{-8}$ at $f\sim 1/(7\ yr)$ according to the original 
analysis \cite{StringBoundMPA}, or $\Omega_{g}h^{2} < 9.3\times 10^{-8}$ 
based on the Bayesian approach \cite{StringBoundMPA1} where $h^{2}\approx 0.65^{2}=0.42$.  
The two constraints lead to $A < 8.25\times 10^{-47}\ Hz^{7/3}$ for the original analysis, 
or $A < 1.28\times 10^{-46}\ Hz^{7/3}$ for the Bayesian approach.  
From the LIGO S4 science run the limit on the isotropic background for the scale-invariant 
case is $\Omega_{g}=1.20\times10^{-4}$ at $f=100\ Hz$, and the limit for the constant 
strain power case is $\Omega_{g}=5.13\times10^{-5}$ at $f=100\ Hz$ 
\cite{OmegaBoundLIGO}.  
From the upper limits we can obtain the upper bounds $A < 1.94\times 10^{-40} \ Hz^{7/3}$ 
and $A < 8.31\times10^{-41}\ Hz^{7/3}$ for the two cases, respectively.  
The result of LIGO science run S5 further constrains the upper limit of frequency independent $\Omega_{g}$ below 
$6.9\times10^{-6}$ at $f=100\ Hz$ \cite{OmegaBoundLIGOS5}, giving the upper bound $1.12\times10^{-41}$ on A.
From these constraints we know that the lowest upper bound is 
$A < 8.25\times 10^{-47}\ Hz^{7/3}$, 
but to prevent the spurious background being selected, the more conservative upper bound 
$A < 1.28\times 10^{-46}\ Hz^{7/3}$ is chosen for the Max(A).  
Therefore the $\delta_{A}$ is set as $1.28\times 10^{-46}\ Hz^{7/3}$.

\subsection{Results}
\begin{figure}
\includegraphics[scale=0.31]{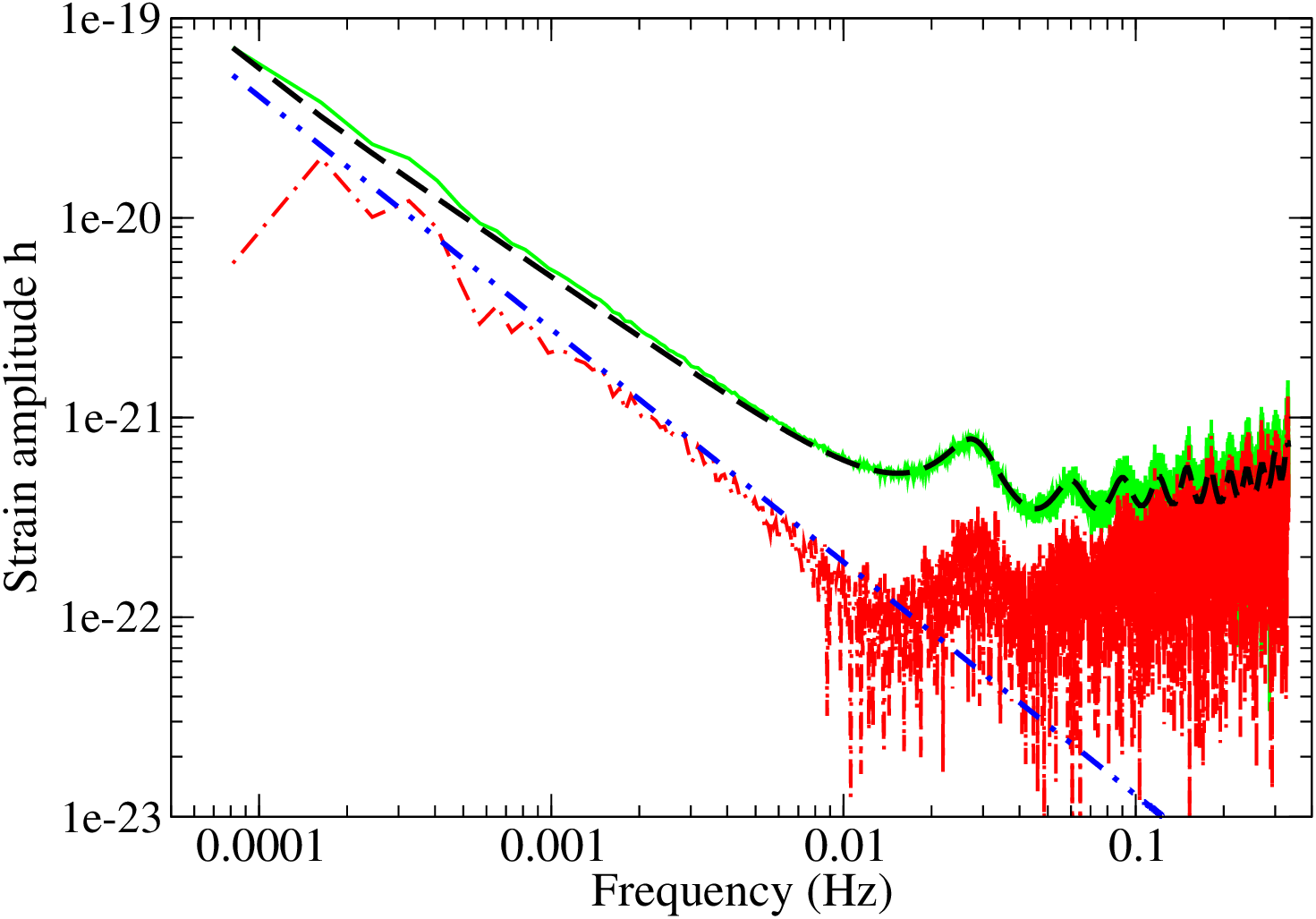}
\caption[The LISA detection threshold for cosmic strings]{
  The green (solid) curve is combined spectrum including both cosmic string signals and 
  instrumental noise.  The red (dash-dot) curve is the residuals after fitting.  
  The blue (double-dot dash) line is the background produced by 
  cosmic strings in the spectrum.  The black (dash) line is the recovery 
  of acceleration noise suggested by fitting with the $M_{0}$ model.  
  The observation duration is 12288 seconds.
  }
\label{fig:string-trial-a}
\end{figure}
Then the existence of the background produced by cosmic strings in the data sets is 
suggested by the model selection method employing the Laplace approximation 
as described in Sec.\ \ref{sec:model-selection}.  
From model selection mothed, the probability ratio of $M_{1}$ to $M_{0}$ is given by
\begin{equation}
\frac{P(M_{1}|I)}{P(M_{0}|I)} = \frac{1}{\delta A}
\frac{\sqrt{Det(\sigma_{1})}}{\sqrt{Det(\sigma_{0})}}
\times\frac{L(\bar{x}_{1}\ |M_{1}, I)}{L(\bar{x}_{0}\ |M_{0}, I)},
\label{eq:string-compare}
\end{equation}
where $L$ is likelihood function, $I$ is background information.  
$\bar{x}_{i}$ and $\sigma_{i}$ are the best estimate and the covariant matrix 
with respect to model $M_{i}$.  
$\delta A$ is the bounded range of parameter $A$.  
$M_{0}$ and $M_{1}$, described by Eq.\ \eqref{eq:M-0} and 
Eq.\ \eqref{eq:power-string+noise} respectively, 
are the model for the cosmic background being absent or present in the spectrum.  
The likelihood ratio $L(\bar{x}_{1}\ |M_{1}, I)/L(\bar{x}_{0}\ |M_{0}, I)$ 
is computed through Eq.\ \eqref{eq:marginalised-likelihood-ratio}.
\begin{table*}
\caption{\textbf{Analysis Result of Gravitational-wave Background by Cosmic Strings.}  
The first column A is 
the real value of the background.  The second column a is the magnitude of random acceleration fitted by the $M_{0}$ model.  
The third column a is the magnitude of random acceleration fitted by the $M_{1}$ model.  $\bar{A}$ in the forth column is the 
estimation of the background strength by the $M_{1}$ model.  The $\rho_{aA}$ is the correlation between a and A in the $M_{1}$ 
model.  R and P are the likelihood and probability ratio of $M_{1}$ to $M_{0}$, respectively. }
\begin{ruledtabular}
\begin{tabular}{ccccccc}
&{ $M_{0}$ Model} & \multicolumn{3}{c}{{ $M_{1}$ Model}} & & \\
{ $A\ (\times 10^{-49}\ Hz^{7/3})$} &{ $a\ (\times 10^{-15}\ m/s^{2})$} & 
{ $a\ (\times 10^{-15}\ m/s^{2})$} & { $\bar{A}\ (\times 10^{-49}\ Hz^{7/3})$} & { $\rho_{aA}$}
& { R} & { P}\\ \hline 
{4.2} & {$2.213\pm 0.006$} & {$2.200\pm 0.007$} & {$2.73\pm 0.56$} & {-0.48} 
& {$4.81\times 10^{4}$} & {20.60}\\
{4.5} & {$2.214\pm 0.006$} & {$2.200\pm 0.007$} & {$3.02\pm 0.56$} & {-0.485}
& {$3.28\times 10^{5}$} & {143.26}\\
{4.8} & {$2.215\pm 0.006$}  & {$2.200\pm 0.007$} & {$3.35\pm 0.55$} & {-0.48}
& {$2.53\times 10^{6}$} & {$1.06\times 10^{3}$}\\
{5.1} & {$2.216\pm 0.006$} & {$2.200\pm 0.007$} & {$3.66\pm 0.55$} & {-0.48} 
& {$2.21\times 10^{7}$} & {$9.28\times 10^{3}$}\\
{5.4} & {$2.217\pm 0.007$} & {$2.200\pm 0.007$} & {$3.98\pm 0.55$} & {-0.48} 
& {$2.30\times 10^{8}$} & {$9.60\times 10^{4}$}\\
\end{tabular}
\label{String-results}
\end{ruledtabular}
\end{table*}
The two models are applied to analyse same data set subsequently.  The best estimates and 
the covariant matrix with respect to each model will be given.  Next, the best estimates 
for the two models are used to compute their likelihood ratio.  Finally, combining the 
likelihood ratio with the determinants of the covariant matrices, 
Eq.\ \eqref{eq:string-compare} provides the Bayes factor of two models.  
If the ratio is higher than 1 and reach 
to certain level, it indicates that the instrumental noise with cosmic string model is the 
better description of data, which is the evidence of cosmic strings existence;  
otherwise, pure instrumental noise will be preferred and the existence of cosmic strings will 
be declined.  
The ratio of 20, which means the existence of cosmic strings is 95\% likelihood, 
is our definition of LISA detection threshold.   

\begin{figure}
\includegraphics[scale=0.31]{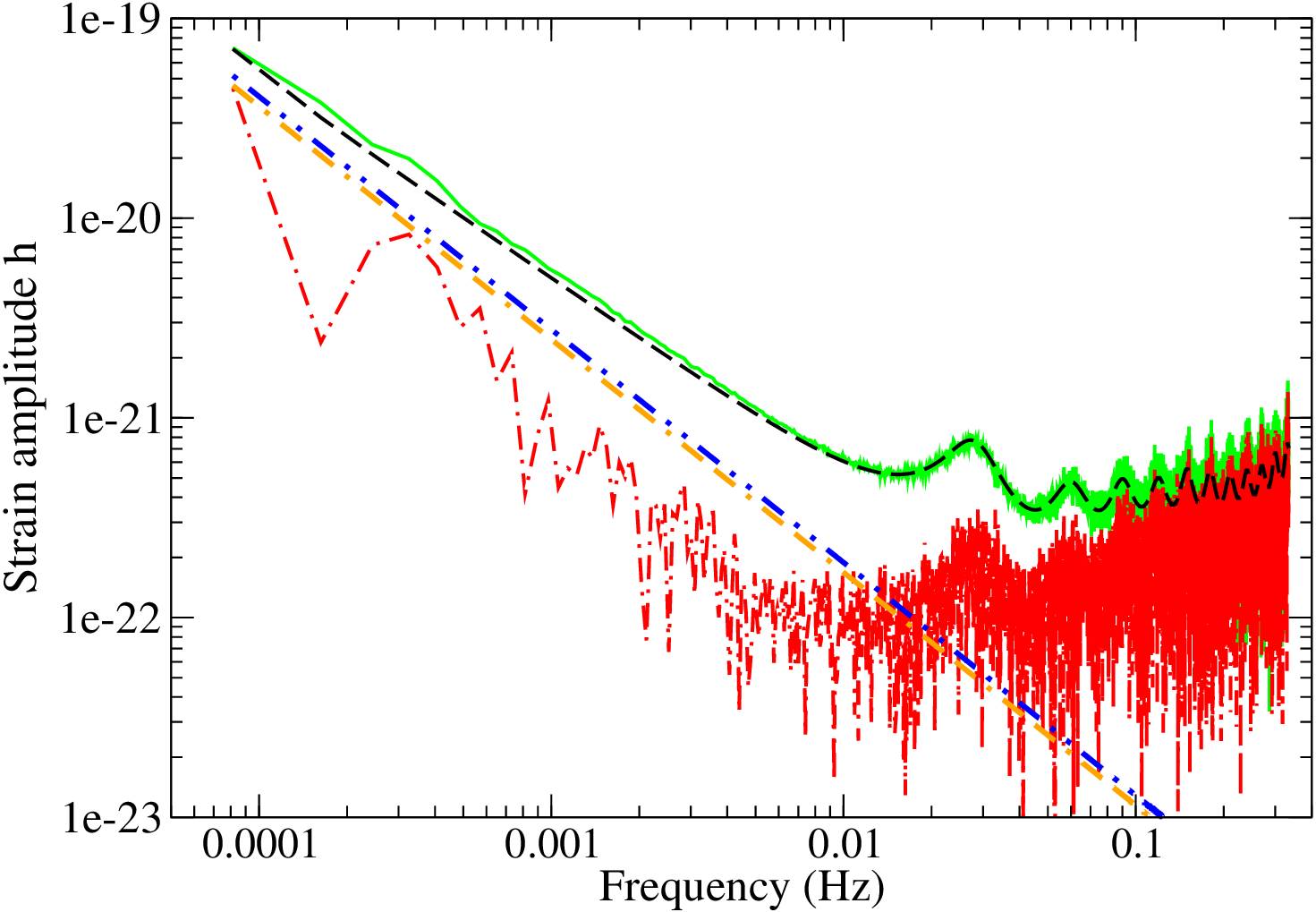}
\caption[The LISA detection threshold for cosmic strings]{
  The green (solid) curve is the full spectrum, and the blue (double-dot dash) line is 
  the background fed into the spectrum, which are the same with those in 
  Fig\ \ref{fig:string-trial-a}.  The black (dash) curve shows the recovered power 
  caused by the random acceleration.  
  The yellow (double-dash dot) line represents the recovered background produced 
  by cosmic strings, which is slightly under the real one.  
  The red (dash-dot) curve is the noise part of the data, representing the difference 
  between the data and the $M_{1}$ model.  
  }
\label{fig:string-trial}
\end{figure}
To pin down the LISA detectability of cosmic string spectrum strength, 
firstly we apply the model selection algorithm in the analysis of a full spectrum containing 
both instrumental noise and a background generated by cosmic strings corresponding to 
a specific strength.  
If the $M_{1}$ model is selected, it means that the strength of the background is stronger 
than the detection threshold.  In this case we decrease the strength of the background to 
generate a new full spectrum, and then re-apply the model selection algorithm in the analysis 
until the $M_{0}$ model is selected.  
On the other hand, if $M_{0}$ model is selected, it means that the strength of the background 
is weaker than the detection threshold.  In the case we enhance the strength of the background 
to generate a new full spectrum, and then re-analyse the data set until the $M_{1}$ model 
is selected.  The strength of the background giving the odds ratio $P(M_{1})/P(M_{0})$ of 
20 is the detection threshold for the gravitational-wave background produced by cosmic 
strings.

\begin{figure}
\includegraphics[scale=0.31]{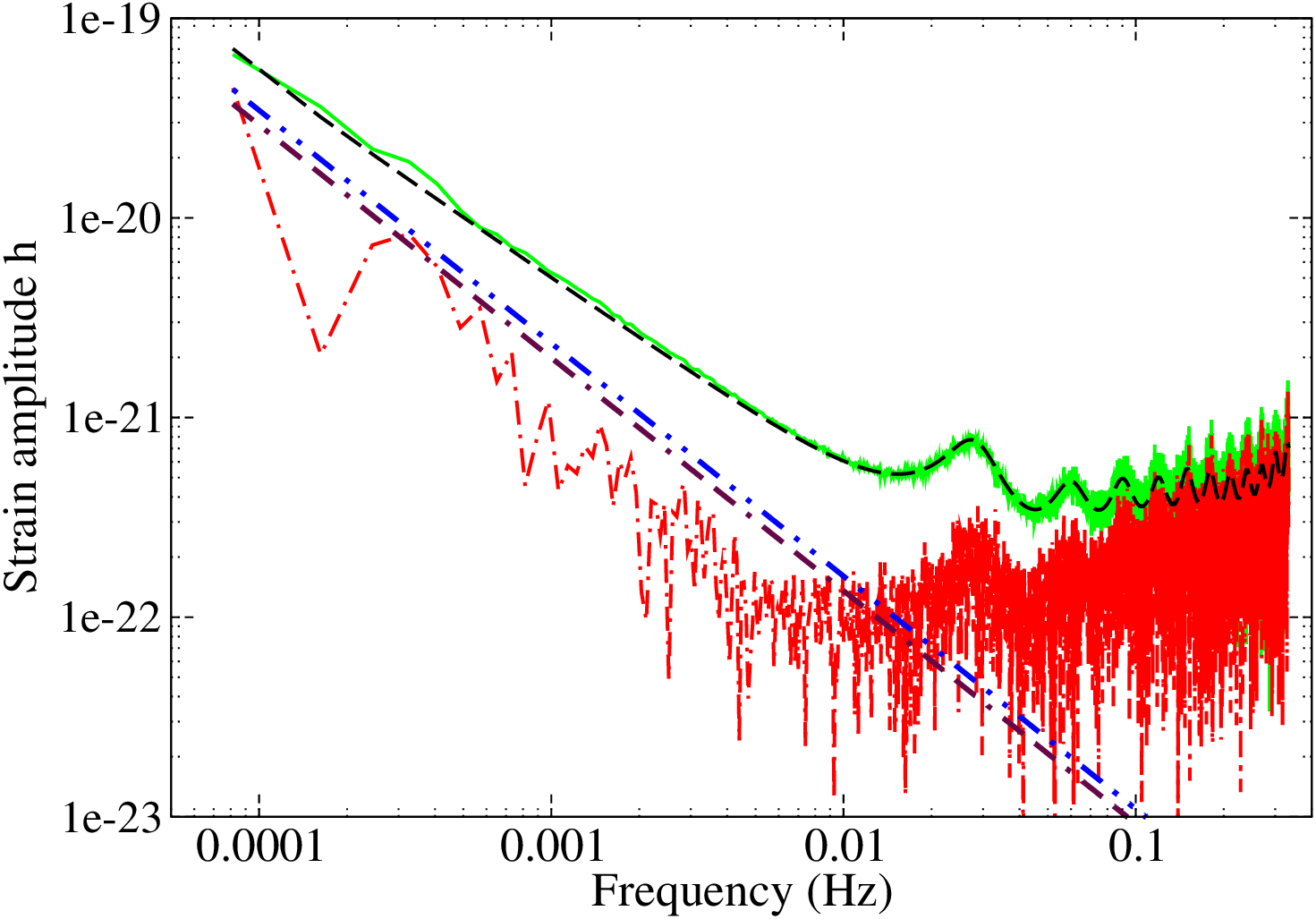}
\caption[The LISA detection threshold for cosmic strings]{
  The green (solid) curve is combined spectrum including both cosmic string signals and 
  instrumental noise.  The red (dash-dot) one is the residuals after fitting.  
  The blue (double-dot dash) line is the threshold for cosmic string detection.  
  The black (dash) and yellow (double-dash dot) lines are the recovery 
  of acceleration noise and string spectrum suggested by fitting with the $M_{1}$ model, 
  respectively.  The observation duration is 12288 seconds.
  }
\label{fig:string-threshold}
\end{figure}

To begin the study of the performance of the algorithm in the analysis of the 
gravitational-wave background produced by cosmic strings we apply the algorithm to 
the full spectrum with the background strength $A=7.7\times 10^{-49}$.  
At first the $M_{0}$ model is used to analyse the data set.  
Since it just contains one parameter $a$, we generate 10 chains to test the convergence.  
From the samples we obtain $a=2.22\times10^{-15}\pm6.79\times10^{-18}\ m/s^{2}$.  
The analysis result is shown in Fig\ \ref{fig:string-trial-a}.  From the figure we can notice that 
the noise character is different from the LISA sensitivity curve below 10 mHz.  
Next, the $M_{1}$ model is used to analyse the same data set.  It includes two parameters, 
$a$ and $A$, so we generate 20 chains to test the convergence.  
From the samples we have $a=2.20\times10^{-15}\pm7.27\times10^{-18}\ m/s^{2}$ and 
$A=(6.11\pm 0.57)\times10^{-49}\ Hz^{7/3}$.  The estimate strength is deviated from the true value 
over 2-$\sigma$ but within 3-$\sigma$.  The correlation between the parameters is -0.48.  
This may be due to the similarity between the formulae for the acceleration and 
the background.  Because the power contained in a frequency bin is fixed, the power 
accounted for the background will be less if that accounted for the acceleration is more, 
and vice versa.  
The Fig\ \ref{fig:string-trial} illustrates the recovered background estimated by the 
$M_{1}$ model.  
The likelihood ratio of $M_{1}$ to $M_{0}$ is $3.99\times 10^{15}$.  The probability 
ratio is $1.67\times 10^{12}$, suggesting that the existence of the background produced by 
cosmic strings is preferred.

In this case the gravitational-wave background is successfully recovered.  
To look for the detection threshold for the background 
we decrease the strength of the background in the full spectra, and apply the model 
selection algorithm to compute the probability ratio of $M_{1}$ to $M_{0}$ model.

As shown in the Table \ref{String-results}, the cosmic string spectrum in the standard 
scenario ($p=\epsilon=1$) reaches the detection threshold if $A=4.2\times10^{-49}$.  
Fig.\ \ref{fig:string-threshold} exhibits the LISA detectability limit for cosmic strings by 
strain amplitude.  If the amplitude of cosmic string spectrum is higher than the blue line, 
the existence of the background is suggested by the algorithm; otherwise, it is disfavored.  
From Eq.\ \eqref{eq:A-value} we obtain that the detection threshold of $G\mu$ is $3.90\times 10^{-11}$ 
in the standard scenario, 
and the detectable region in parameter space is defined by 
$G\mu^{2/3}/p\epsilon^{1/3}>1.15\times 10^{-7}$.  

\bibliography{String_analysis}

\end{document}